# Transfer and Online Reinforcement Learning in STT-MRAM Based Embedded Systems for Autonomous Drones

Insik Yoon[1], Aqeel Anwar[1], Titash Rakshit[2], Arijit Raychowdhury[1]

[1] Georgia Institute of Technology, Atlanta GA, USA
[2] Samsung semiconductor, advanced logic lab, Austin TX, USA
Email: {iyoon, aqeel.anwar}@gatech.edu, titash.r@samsung.com, arijit.raychowdhury@ece.gatech.edu

***Abstract*—In this paper we present an algorithm-hardware co-design for camera-based autonomous flight in small drones. We show that the large write-latency and write-energy for non-volatile memory (NVM) based embedded systems makes them unsuitable for real-time reinforcement learning (RL). We address this by performing transfer learning (TL) on meta-environments and RL on the last few layers of a deep convolutional network. While the NVM stores the meta-model from TL, an on-die SRAM stores the weights of the last few layers. Thus all the real-time updates via RL are carried out on the SRAM arrays. This provides us with a practical platform with comparable performance as end-to-end RL and 83.4% lower energy per image frame.**

## I. INTRODUCTION

Over the past decade, there has been considerable success in using Unmanned Aerial Vehicles (UAVs) or drones in varied applications such as reconnaissance, surveying, rescuing and mapping. Irrespective of the application, navigating autonomously, particularly with camera based inputs, is one of the key desirable features for small drones, both indoors and outdoors. In recent years, reinforcement learning (RL) has been extensively explored for different type of robotic tasks, including drone navigation and collision avoidance. RL, in spite of its bio-mimetic approach, is computationally challenging [1,2]. The agent (drone) needs to collect visual data and train a neural network based model in real-time [2,3]. For a given velocity of the drone, the corresponding distance traveled between two frames ($d_{frame}$), and the minimum distance between obstacles (a measure of clutter in the environment), we can calculate the minimum number of frames/second (fps) required for collision avoidance (summarized in Fig. 1). Since the drone needs to train on acquired data at least at the same rate as the fps, the amount of computation that needs to be performed is prohibitively large for embedded systems that can be mounted on small drones. Further, the emergence of non-volatile memory (NVM) [4-6] technologies that exhibit high-density and low-standby-power aims to disrupt the design of embedded systems. In spite of their advantages, all NVM technologies shows high write latency and energy. This makes them unsuitable for storing model weights in real-time RL systems such as drones, both in terms of meeting an fps (or, velocity) requirement and energy target.

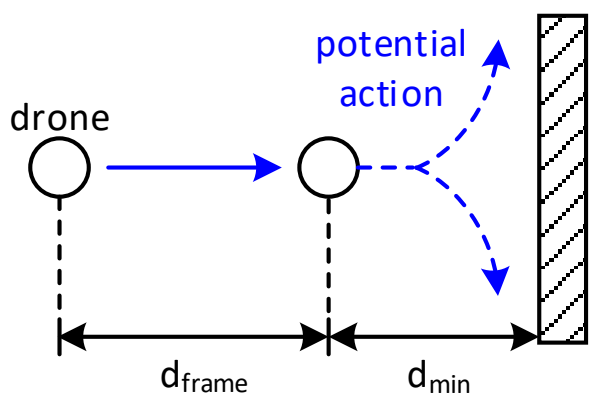

(a)

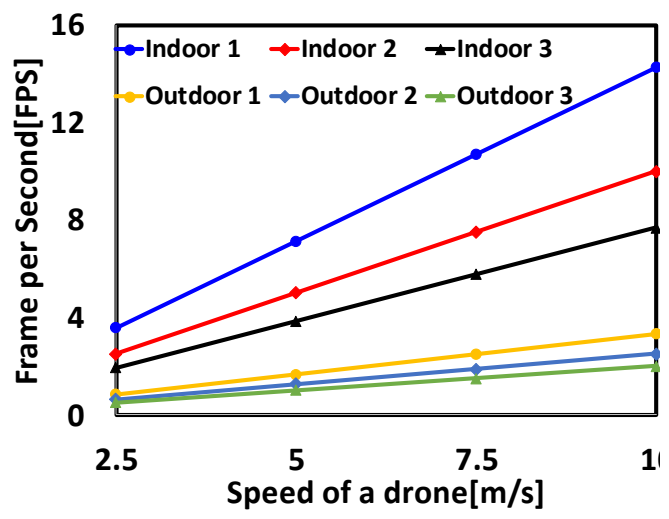

(b)

| Environment | $d_{min}$[m] |
|---|---|
| Indoor 1 | 0.7 |
| Indoor 2 | 1 |
| Indoor 3 | 1.3 |
| Outdoor 1 | 3 |
| Outdoor 2 | 4 |
| Outdoor 3 | 5 |

| | Image frame per second[FPS] | | | | | |
|---|---|---|---|---|---|---|
| $v_{drone}$[m/s] | Indoor 1 | Indoor 2 | Indoor 3 | Outdoor 1 | Outdoor 2 | Outdoor 3 |
| **2.5** | 3.571 | 2.5 | 1.923 | 0.833 | 0.625 | 0.5 |
| **5** | 7.142 | 5 | 3.846 | 1.666 | 1.25 | 1 |
| **7.5** | 10.71 | 7.5 | 5.769 | 2.5 | 1.875 | 1.5 |
| **10** | 14.28 | 10 | 7.692 | 3.333 | 2.5 | 2 |

(c)

Fig.1. (a) Definition of minimum distance required for obstacle avoidance ($d_{min}$). $d_{frame}$ = distance that drone moves between frames. (b) Frame per second vs. speed of a drone for sample indoor and outdoor environments (c) $d_{min}$ setting for different environment and minimum FPS needed for obstacle avoidance for different environments

To address this fundamental challenge, we propose an algorithm-hardware co-design where we show:

1. Context-aware transfer-learning (TL) augmented with RL. During TL phase, before deployment, *a drone is trained in complex meta-training-environments* (indoor and outdoor). This is accomplished via reinforcement learning (RL) on the meta-training-environments.
2. *At the time of deployment, the correct meta-model (indoor or outdoor model) obtained from TL is downloaded to the drone* whose embedded platform consists of a large, stacked-NVM array and a smaller (~30 MB) on-die SRAM. As a part of this study, we consider spin-transfer-torque (STT-RAM) as the NVM of choice. A part of the model (last few layers of the neural network) are stored in the on-die SRAM.
3. After deployment, *the drone performs real-time RL*; but instead of learning all the model parameters, it only trains the last few layers which are stored in the SRAM. This results in only read accesses from the NVM array during flight (inference/ forward propagation of data) and all the necessary write operations are executed on the on-die SRAM. Since the coarse features of the environment (obtained from TL) are stored in the first several layers of the network, the proposed algorithm works successfully as the drone needs to learn only the environment specific finer features (online RL) in real-time.

We show that the proposed TL followed by environment-specific RL over the last few layers achieves comparable accuracy as E2E RL. While E2E RL on an environment is not feasible with NVM based embedded platforms (in terms of latency and energy requirements), our proposed solution archives real-time operation with 79.4% (83.45%) decrease in latency (energy) compared to a baseline E2E RL system.

## II. REINFORCEMENT LEARNING FOR DRONE NAVIGATION

### A. Basics of End-to-End Reinforcement Learning

The idea of Reinforcement Learning (RL)[1] is to learn a control policy by interacting with the environment. In supervised learning we have access to the labelled data. On the other hand, we don't have a-priori access to the labeled data in RL; rather, the agent continuously interacts with environments (state space), takes actions in the space and updates the functional mapping between the state and action spaces. In RL, when the agent is placed in a new environment, its initial actions are random. With every action taken, the agent is presented with a reward. This reward mimics the high-level goal that we want the system to achieve. The objective for the agent is to learn a policy that maximizes the long-term reward. At time step $t$, the agent senses the current state of the system $s_t$. With every action taken, the agent moves in the environment and observes a new state $s_{t+1}$. This new state along with the previous state is used to evaluate a reward $r_t$ for the action taken. The goal of RL is to determine subsequent actions such that the long-term discounted return $R_t = \sum_{i=t}^{T} \gamma^{i-t} r_i$ (where, $\gamma$ is the discount factor) is maximized. This maximization is done by the use of the Bellman Equation on the data tuple ($s_t$, $a_t$, $s_{t+1}$, $r_t$). In the Q-learning RL algorithm [1,2] each state-action pair is assigned a $Q$ value, $Q(s,a)$. The $Q$ value signifies how favorable an action, $a$ is given the state, $s$. As the agent trains itself, the $Q$ values are updated based on the reward $r$ as:

$$Q(s,a) = r + \gamma \max_{a'} Q(s', a') \quad (1)$$

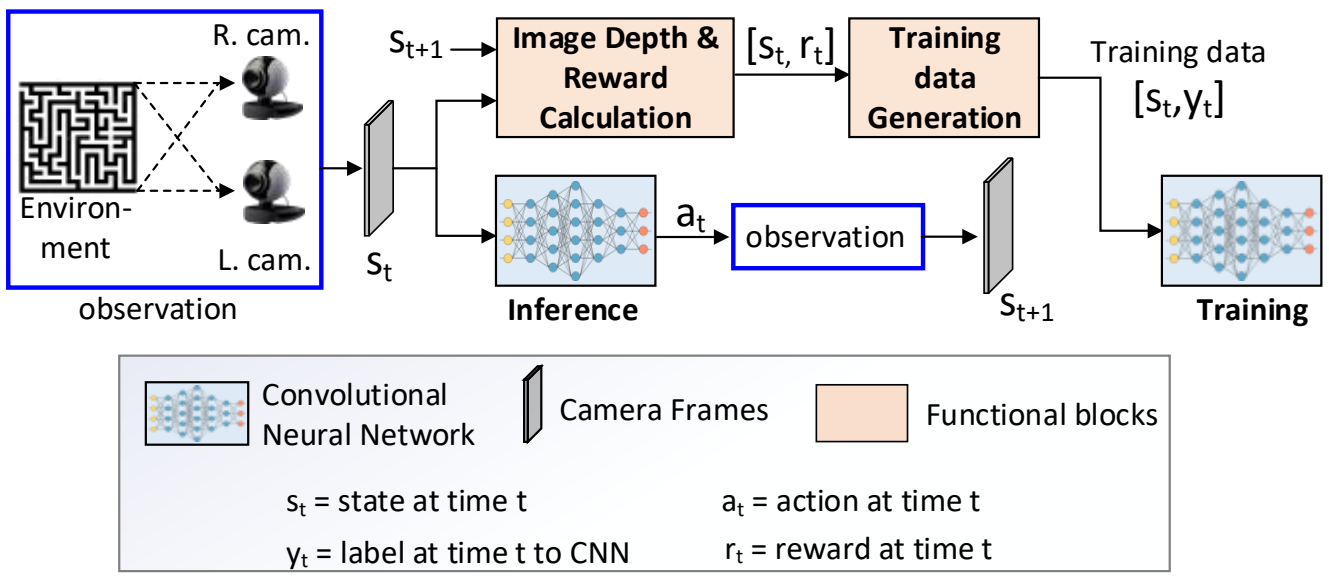

Fig.2. Reinforcement Learning (RL) network architecture

The agent selects an action, $a_t = \max_{a'} Q(s_t, a')$ and consequently maximizes the discounted return in the long run.

*B. RL in Camera Based Navigation in Drones*

The problem at hand is end-to-end navigation via collision avoidance (long term goal) in drones using a camera system. We map the navigation problem to the RL problem as follows. The state at time instant $t$, $s_t \in S$ is the output of the camera and hence is an image. At any given state, we can take any action $a_t \in A$ where $A$ is the action space. We have limited the action space to five values $A = \{0,1,2,3,4\}$ where under the action 0 the drone moves forward, 1 and 3 the drone turns left with turn angles $25^O$ and $55^O$ respectively and 2 and 4 the drone turns right with turn angles $25^O$ and $55^O$. These five actions are sufficient for the drone to navigate in its surrounding. We used the disparity map from stereo camera to generate an approximate depth map of the camera frame [2]. We use a part of the depth map towards reward generation in a manner described in [3]. The depth map generated is segmented into a smaller window in the center. The reward is taken to be the average depth in this center window. The closer the drone is to the obstacles, the lesser the average depth in the center window and the smaller the reward is. A deep Convolutional Neural Network (CNN) is used to estimate the $Q$ values for the states. The input to the CNN is the resized camera frame $s_t \in \mathbb{R}^{nxn}$ where $n = 224$. The network architecture is based on a modified Alexnet model [9]. The network consists of 5 convolutional layers and 5 fully-connected layers, optimized for autonomous navigation. The network architecture and hyper-parameters are shown in Figure 3. As the network trains during flight, it continually learns the weights of the model and presents a continuously improving functional mapping between the state and the action.

*C. Challenges of End-to-End (E2E) RL in Embedded Systems*

In a true biologically-inspired system, an autonomous drone should learn to navigate via E2E RL [3]. It should start from a random initialization of model weights and learn the final model iteratively via interactions with the environment. Although feasible [3], this faces two fundamental challenges:
1. During exploration, the drone will take random, often incorrect actions and collide with obstacles. These unsafe actions can cause damage to the drone or the environment.
2. Further, E2E RL is computationally extremely challenging. It is impossible to achieve autonomy via RL in small form factor drones, without additional off-board infrastructure [3]. As we move into an era of powerful edge-nodes, the computing architectures are becoming capable of supporting large CNN models in-situ. However, for high density and low stand-by power non-volatile memory (NVM) is emerging. STT-MRAM is becoming a mature NVM technology, and in-spite of its high-density, endurance, nano-second read speeds, the process of write in STT-MRAMs is expensive both in-terms of energy and latency. This makes it practically impossible to use STT-MRAM for model storage in RL

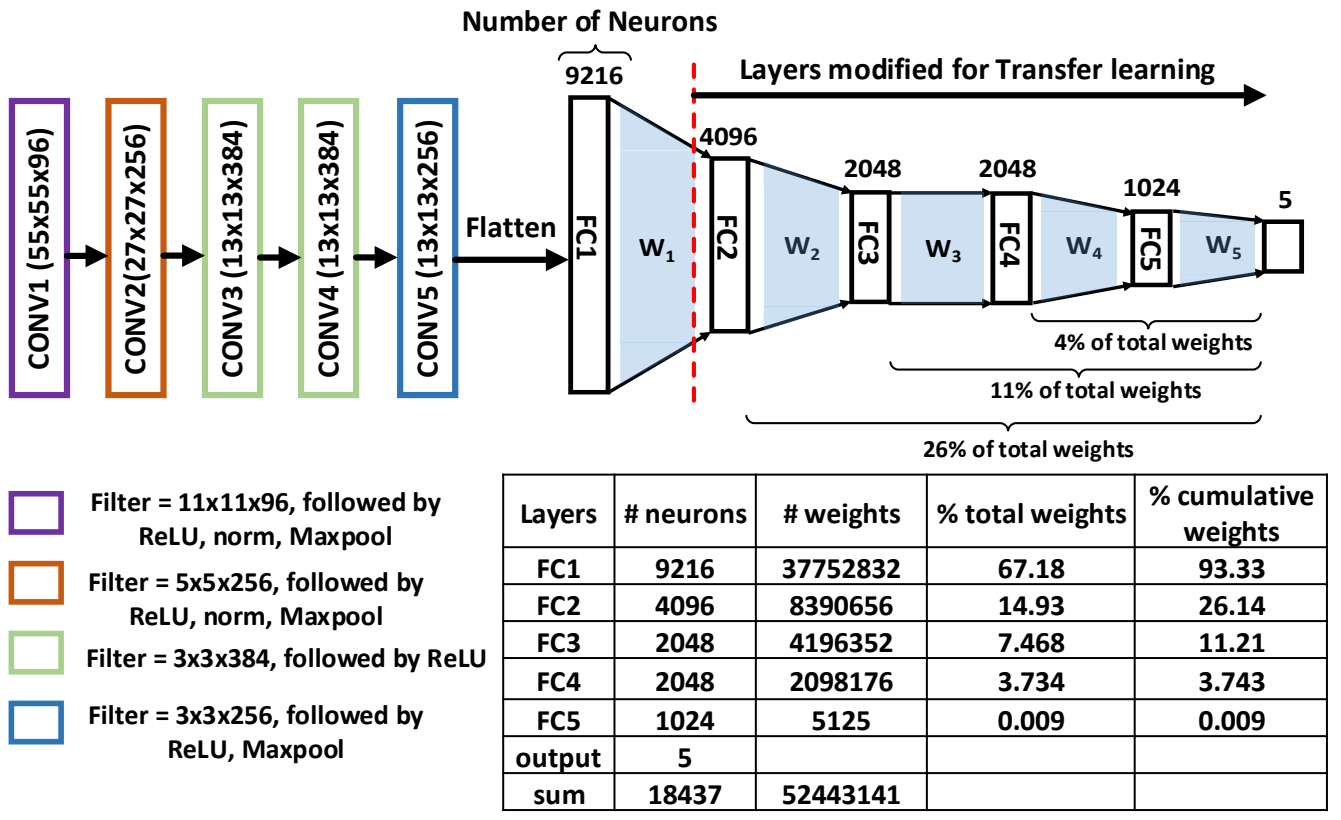

| Layers | # neurons | # weights | % total weights | % cumulative weights |
|---|---|---|---|---|
| FC1 | 9216 | 37752832 | 67.18 | 93.33 |
| FC2 | 4096 | 8390656 | 14.93 | 26.14 |
| FC3 | 2048 | 4196352 | 7.468 | 11.21 |
| FC4 | 2048 | 2098176 | 3.734 | 3.743 |
| FC5 | 1024 | 5125 | 0.009 | 0.009 |
| output | 5 | | | |
| sum | 18437 | 52443141 | | |

(a)

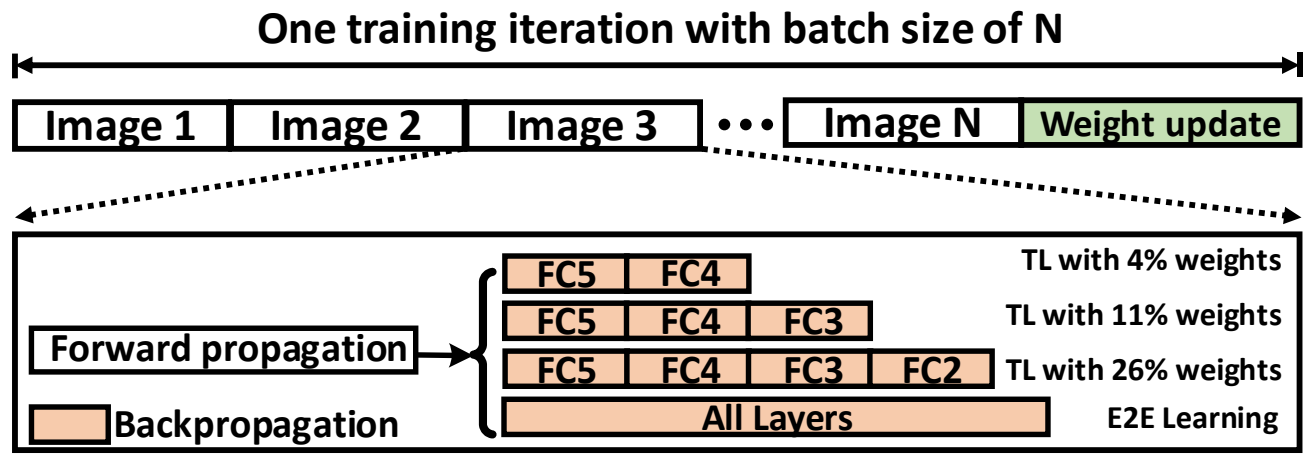

(b)

Fig. 3. (a) Modified AlexNET [9] for the proposed system (b) 3 configurations where 4,11 and 26% weights are learnt in real-time. This is in contrast to E2E RL, where the entire network is learnt in real-time.

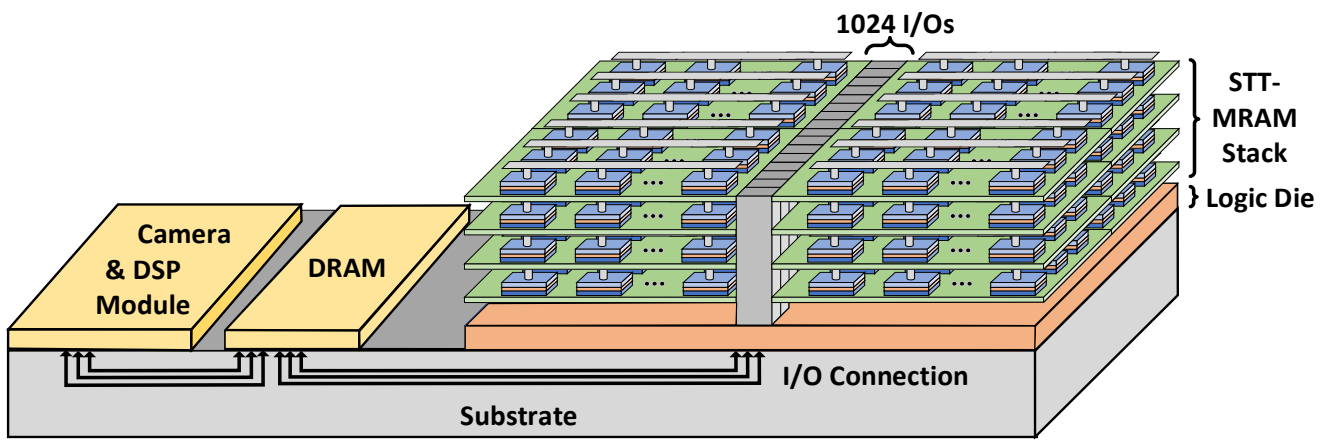

(a)

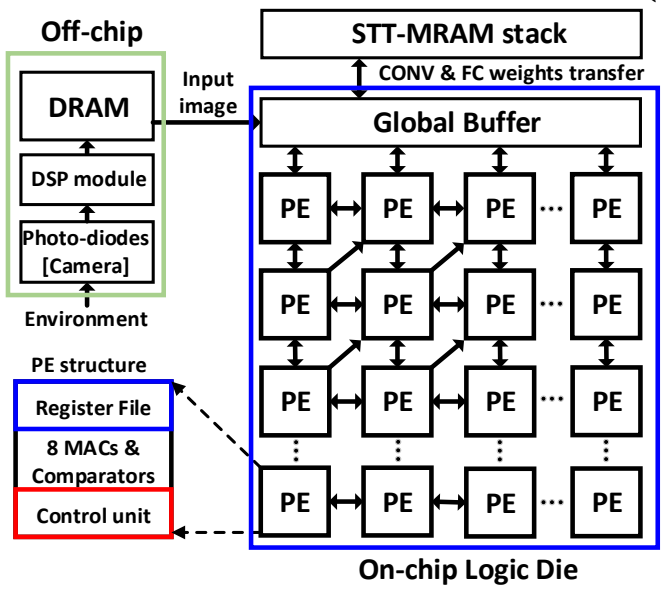

| Technology | NanGate 15nm FreePDK |
|---|---|
| Number of PEs | 1024 (32 row, 32 column) |
| Global buffer /scratchpad | 30MB/4.2MB |
| Register File per PE | 4.5KB |
| Operation voltage | 0.8V |
| Clock speed | 1Ghz |
| Peak Throughput | 1.5TOPS/W |
| Arithmetic precision | 16 bit fixed-point |
| Bandwidth between PEs | 128 bit |

Fig.4. (a) 3D view of the hardware platform (b) System architecture and parameters as extracted post-synthesis in 15nm nangate PDK.

systems, where every action needs a corresponding update of the entire model via backpropagation and gradient descent.

### *D. Transfer Learning(TL) with Real-Time RL*

To address the challenges of E2E RL mentioned above, we propose transfer learning with real-time RL– an algorithm-hardware co-design that matches the learning algorithm with a hierarchical memory sub-system that we describe below. Transfer learning (TL) is a technique where a model trained on a meta-task and is transferred to an agent to minimize the need for online real-time RL and reduce unsafe actions early on [7, 8]. In our proposed system the agent learns on an embedded platform in the following steps:

1. The CNN is first trained on a meta-environment and is downloaded (on NVM as well as on-die SRAM) as a TL model prior to deployment. We use two types of simulated training environments (indoor and outdoor), although it can be extended to other environment types as well.
2. The downloaded TL model is then trained in real-time using RL. However during real-time learning, we only update the last few fully connected (FC) layers of the model, which resides on an embedded SRAM array. This allows us to use the stacked- NVM array for read (R)-only (for inference) and on-chip SRAM arrays for read and write (W) (for learning). We show that TL followed by RL on the last few layers performs equally well as compared to E2E RL, at a significant reduction in energy/training and latency, which finally improves the drone's battery life and speed (Fig. 1).

We study three different embedded architectures with different on-die SRAM capacity – capable of storing 26% (FC2+FC3+FC4+FC5), 11% (FC3+FC4+FC5) and 4% (FC4+FC5) of the total weights of the network. Fig. 3(b) describes the procedure for on-line training. One training iteration with batch size of *N* images is defined as the sum of *N* iterations of forward & backpropagation with one image. Based on the TL configurations, we back-propagate last 2/3/4 layers of the network. In E2E learning (baseline), we back-propagate across all the layers, as shown in Fig. 3(b).

## III. PROPOSED SYSTEM ARCHITECTURE

Our system architecture, which includes a systolic, array-processor [9] with on-die SRAM (buffer memory) and stacked STT-MRAM arrays (Fig 4). We use the high-bandwidth-memory (HBM) architecture for STT-MRAM and borrow the organization of the sub-arrays and the local/global IO from JEDEC [10]. The DRAM arrays of traditional HBM are replaced by STT-MRAM providing a realistic and emerging platform for an embedded system with high-bandwidth IO, based on [10]. A camera system (with the necessary pre-processing DSP) and a DRAM-based buffer memory is shown in Fig. 4(a), is integrated on a substrate (which can be a silicon interposer or a package substrate). The camera buffer is connected to the logic die using a DDR6 link.

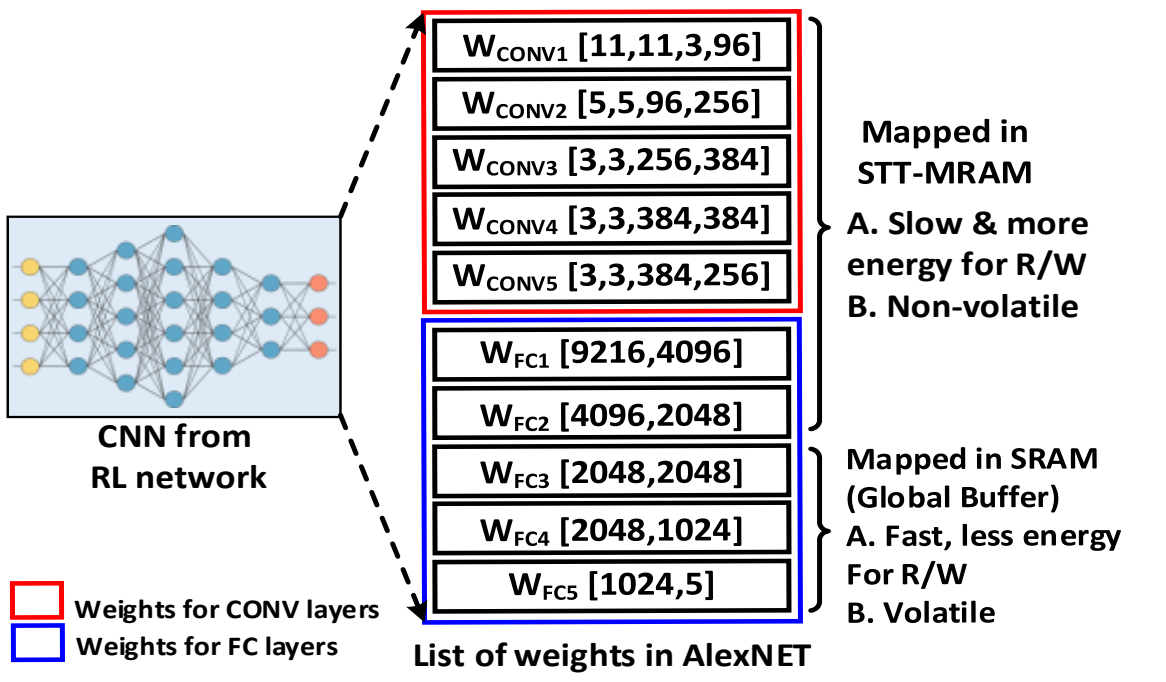

Fig 5. Mapping the weights of the proposed CNN (modified AlexNET) to stacked-STT-MRAM and on-die SRAM in the system

### *A. Off-chip to On-chip Data Movement*

The camera with a DSP module and buffer-DRAM are located off-chip on a shared substrate. The logic die loads one image frame at a time to an on-chip global buffer for taking action and performing RL. In the proposed system, the data flow between DRAM and logic die uses the DDR6 protocol.

### *B. On-chip System Architecture with Stacked STT-MRAM*

A 3D-stacked STT-MRAM [5,6] is stacked on the logic-die in the same way as HBM is currently stacked and the logic die lies at the bottom on the common substrate [10]. The weights of each layers of the network are stored in the STT-MRAM stacks. The number of PEs in the systolic array is 1024 (32 x 32) and each PE has 128 bit connections with 4 nearby PEs and diagonal connections with an upper right PE [9,13]. The global buffer has 4096 connections with 32 PEs in the first row and can broadcast the same data to each row of the PEs. 1024 I/O connections exist between STT-MRAM stack and global Buffer and bandwidth of each I/O is 2Gbit/s [10]. Each PE has a register file, 8 MACs for convolution and vector-matrix multiplication and 8 comparators for rectified linear and maxpool operations. Fig. 4 (b) shows a complete list of system parameters. The whole system is designed, synthesized and in the 15nm nangate technology [15]. All results discussed here are post-synthesis.

| Write latency | Read latency | Write energy* | Read energy* |
|---|---|---|---|
| 30ns | 10ns | 4.5pJ/bit | 0.7pJ/bit |

Table 1. STT-MRAM parameters used in the system [4][5][6]
*write/read energy includes energy of IO, peripheral and STT-MRAM array

### *C. Why STT-MRAM?*

It is well understood that next-generation memory-intensive learning-based systems require a memory technology which shows high-density, low-standby power (hence NVM) and acceptable R/W speeds. Compared to other NVMs such as Phase-change memory or resistive RAM, STT-MRAM exhibits better read/write latency [12, 13] and is more mature than Ferroelectric FET based RAMs. Further, RRAMs show large device-to-device and cycle-to-cycle variations making it hard to commercialize [11]. Although our study investigates STT-MRAM based stacks, all NVM suffer from high write latency and energy; and hence the algorithm-hardware co-design that we propose is applicable to similar other platforms. The STT-MRAM model parameters are summarized in Table 1.

### *D. Mapping the CNN Model to the Memory System*

Fig. 5 presents the weight mapping of the CNN to the memory system comprising of stacked-STT-MRAM and on-die SRAM. Since we update weight parameters for last 2/3/4 layers of fully connected layers for transfer learning, it is ideal to have enough SRAM-based on-chip global buffer to store weights that need to be updated in real-time. The size of model of the second fully connected layer, FC2, is 29.38MB (each weight parameter is 16 bit fixed point). Therefore, in the proposed design, we store the weights from last three layers only in the global buffer and the cumulative sum of these weights is 12.6 MB. The rest of model weight which consists of the CONV layers and FC1 and FC2 add up to 100MB and are stored in stacked-STT-MRAM array. Further, for weight update in TL, we store the sum of weight and bias gradients

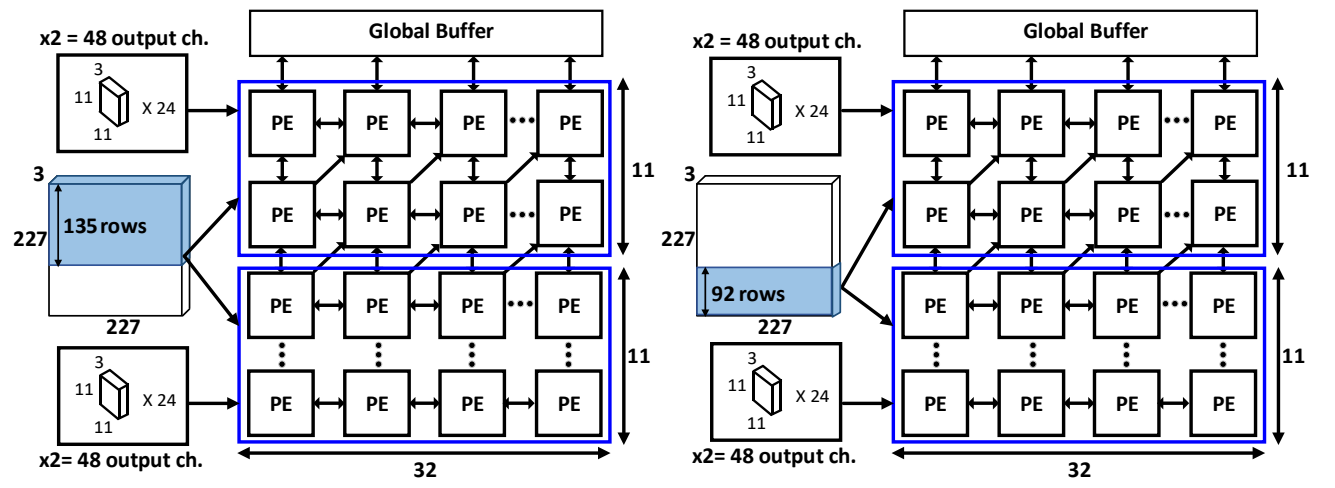

(a) Type I mapping used in CONV 1

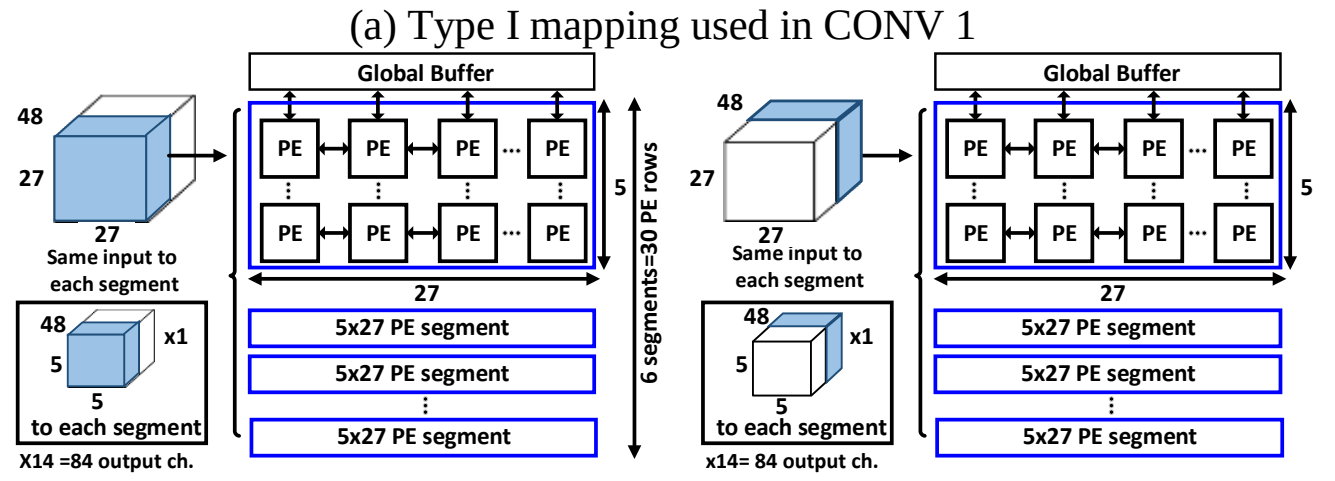

(b) Type II mapping used in CONV 2

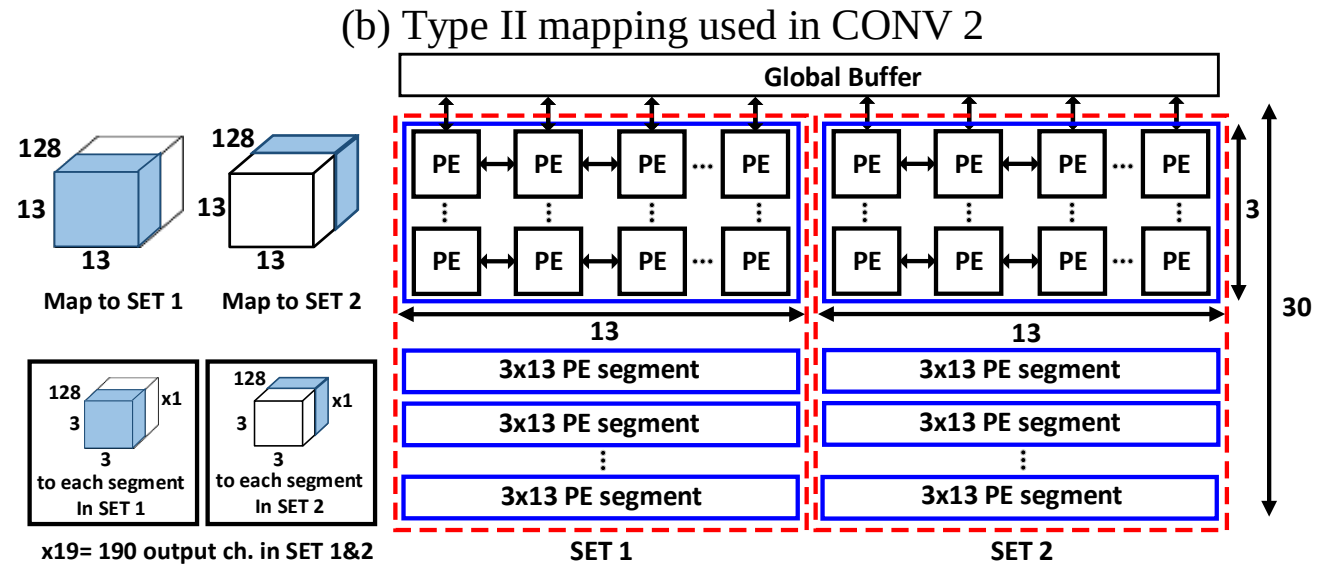

(c) Type III mapping used in CONV 3, CONV 4 and CONV 5

Fig.6: Strategies for mapping weights and data for processing the convolutional layers.

of last 3 layers of the network to the global buffer. Once we have the sum of gradients of weights and bias after processing a batch size of *N*, we need to update the weights as shown in a manner shown in Fig. 3(b). For these scratch-pad calculation, we estimate an additional 12.6 MB of global-buffer. In summary, the global buffer uses 25.2 MB of space to store weights in the last three layers for forward propagation and the sum of weights and bias gradients from the last three layers used in backpropagation. Finally, an added 4.2 MB of global buffer is used as a scratchpad for loading input/weight parameters to PE array and storing intermediate results from PE array. This leads to a total on-chip SRAM size of 29.4 MB, which is at-par with the on-die SRAM capacity of practical embedded systems.

## IV. Forward Propagation Through the CNN

### *A. Forward Propagation in Convolution (CONV) Layers*

Row stationary dataflow architecture is used in the systolic array for convolution in forward propagation [14]. The basic steps are:

1. Input images to the convolution layer are loaded from the global buffer to the local register file (RF) in each PE. We use the diagonal connection to nearby PEs to maximize data reuse within PE array and reduce data movement.
2. Each row of filter weights is broadcasted from the global buffer to the RF in each PE in the same row of the PE array.
3. Row-wise convolution is conducted in the MAC units in each PE and we write the result (*pSUM* ) in the RF.
4. We accumulate the *pSUM* from each PE vertically to the first row of PE arrays and write the convolution results back to the global buffer.

Depending on the height of filter in each CONV layer, we partition the PE array into segments to complete the convolution operation. For example, Fig.6 (a) shows the partition of PE array for the first convolution layer where the filter size is (11, 11, 3, 96) with a stride of 4. The PE array is partitioned into two segment and each segment contains 11x32 PEs. The height of segments is equal to the height of the filter. This is due to the fact that each row of filter is mapped to each row of PE array for row-stationary dataflow. The size of RF inside the PE, the dimension of PE array and filter size of convolution layers determine the mapping scheme of filter and input data to the system. Fig. 6 presents three types of data mapping schemes used in the design. We use Type I on the first convolution (CONV1) layer is shown in Fig.6 (a). Since there are 3 input channels of image and filters in CONV1, an RF size of 4.5 KB is large enough to store each row of filter and image with all the input channel. The same image data is loaded to two segments of the PEs and filters with 24 different output channels are mapped to each segment. Depending on the RF size, the number of output channels of the filters can vary. The number of columns inside the segments determines how many row of images, the system can convolve per cycle. Since we have 32 columns, the system can produce the convolution results of 135 rows of input image in a single cycle. (135 = 32*stride + filter height). Fig.6 (b) presents the TYPE II mapping scheme of data for CONV2. In this case, the number of input channels of filter and input to CONV2 are too large to fit in register file of a PE. TYPE II divides input channels of filter and input into two parts and loads them to PE array. Since the filter height of CONV2 is 5, the PE array is partitioned into 6 segments where each segment dimension is 5x27. Instead of using all 32 columns of PE, 27 columns are utilized because each column generates one row of convolution output. The same image data is mapped to all 6 segments and each segments are mapped with different corresponding filters and each segments generate distinct outputs after computation. Fig.6 (c) presents the TYPE III mapping scheme of data for CONV3. The main

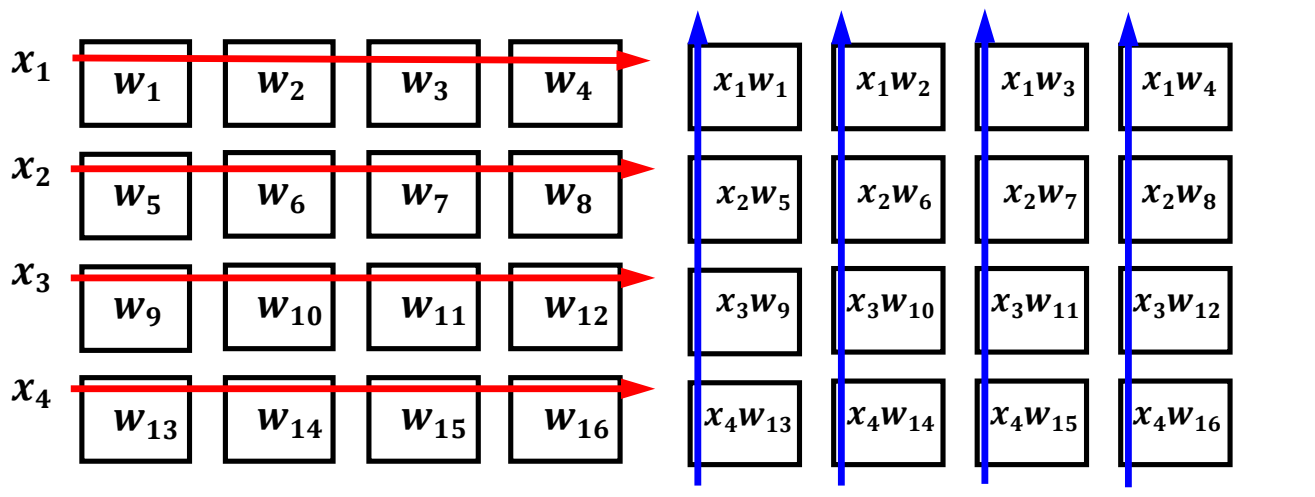

(a) (b)

Fig.7: (a) Row-wise vector propagation in PE array for calculating pSUM (b) Vertical pSUM accumulation for vector-matrix multiplication in forward propagation of FC layers

difference between TYPE II and TYPE III mapping is the existence of set. What we define a set is a collection of PE segments. Since the filter width and height are decreased from CONV2 to CONV3, we can map 2 sets of 10 segments, each segment dimension is 3 by 10 PE, to PE array for CONV3. In TYPE III mapping, the segment size of PE is 3x13 because the filter dimension is (3,3) and the stride is 1. Because the dimension of the segments is lower, we partition the PE array into 2 sets of 10 segments (total 30x26 PE array). Due to the high number of input channels of input and filter to CONV3, we split the input channel of filter and inputs into two parts. Unlike TYPE II, the two parts of input and filter are mapped to each set of the PE array, which enables us to map the input and the filter with all the input channels. After completing *pSUM* in step 4, the convolution results in the first row of set 2 must be transferred to the first row of set 1. For example, the output from PE at 14th column (PE in the 1st column in set 2) must be transferred to the PE in the 1st column in set 1. Then two results from set 1 and set 2 is added together to complete convolution. Since the filter height and width (3,3) in CONV4 and 5 are same as the filter height and width in CONV3, TYPE III mapping scheme is used for CONV4 and 5 as well.

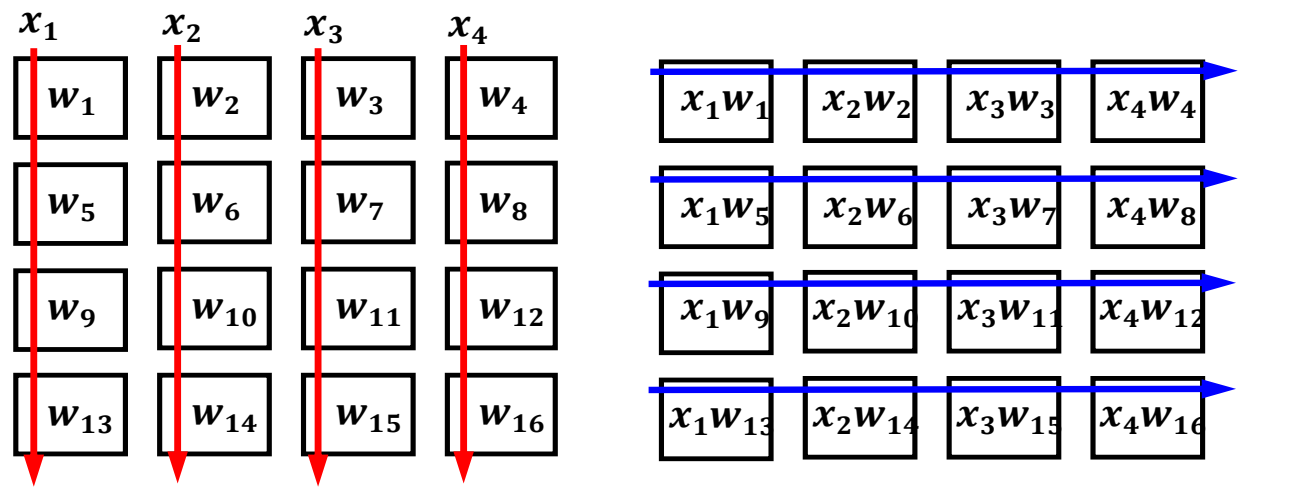

Fig.8: (a) Column-wise vector propagation in PE array for calculating pSUM (b) Row-wise pSUM accumulation for vector-transposed matrix multiplication in backpropagation of FC layers

### B. Forward Propagation in Fully Connected (FC) Layers

In forward propagation through FC layers, vector-matrix multiplication is the primary computation. Fig.7 describes the core operations of PE array for vector-matrix multiplication. After loading matrix components to the PE arrays, the vector elements are propagated row-wise in the PE array and we perform multiplication in each PE. Once the *pSUMs* are generated in each array, they are accumulated vertically and transferred to the global buffer.

## V. Backpropagation and Gradient Descent

For TL followed by online RL, we train last 2/3/4 FC layers of the network. Backpropagation consists of two major computational steps: finding gradients of weights and their biases. Since we use our system to serially process one image at a time for training, the system must store the sum of weight and bias gradient of each image in the global buffer.

### A. Backpropagation architecture of Fully-Connected Layer

The gradient of the weight is the result of multiplication of every vector element in a layer of neurons and every vector element in the gradient of the loss function computed with respect to the neurons in previous layer. Since there is no *pSUM* accumulation involved in calculating bias gradients, the results of multiplication of each PE are directly transferred to global buffer. The gradient of the bias in an FC layer is calculated my multiplying the vector of the gradient of Loss with respect to neurons in previous layer and the transposed weight matrix. The structure of the systolic array enables vector-transposed matrix multiplication without transposing the matrix itself, in a manner describe in [14] Fig. 8 describes the structure of vector-transposed matrix multiplication in the PE array. The vector elements are propagated downwards in each column of the array and the *pSUM* from each PE are accumulated row-wise. The computation is complete when PEs in the last column transfer their results to the global buffer.

### B. Backpropagation architecture of CONV

The backpropagation of CONV layers only happen when evaluating the E2E RL in the system, which is our baseline design. For comparison to the baseline, we benchmark the backpropagation architecture for the entire network. For CONV layers, we use GEMM [16], where the system first reads the data from the STT-MRAM array to the logic die, and expands the inputs to each CONV layers in a 2D matrix. Once the expansion is complete, the backpropagation of CONV becomes same as the backpropagation of FC layers. After the weights of the CONV layers are updated, we write the weights back to the STT-MRAM array. We account for the additional on-chip SRAM requirement for storing the results of the intermediate compute steps.

## VI. Simulation Setup and Results

### A. Hardware Architecture simulation

We used NanGate 15nm FreePDK cell library to evaluate the hardware system performance [15]. We have performed synthesis and place-and-route of the entire system and the results cited here (along with Fig. 4) are post-synthesis.

### B. Simulation on Drone based system

The algorithm is tested on a simulated environment with the dynamics of realistic drones. Simulations were carried out on two types of simulated environments, Indoor and Outdoor. For each of the two categories, complex meta-environments and separate test environments were designed to train and test the performance of the proposed methodology respectively. We used the Unreal Engine 4, used for video game development to design the simulation environments and emulate the necessary physics. This engine interfaces with Tensorflow to train a drone via TL and RL. The web-link for the suite of the environments, videos and corresponding data sets can be found here:<*to be added in the final manuscript*> and the details are beyond the scope of this paper. Typical screen-shots are shown in Fig. 9. The drone is trained in the meta-environment for 60K iterations, initialized with ImageNet weights. The trained weights are then used as initial weights for RL in the respective test environments. For RL, we use 4 topologies, E2E (end-to-end RL) and L2, L3, and L4, where $L_i$ represents TL followed by RL where the last i-layers are trained online. Fig. 10 reports the results for these test environments in terms of cumulative rewards and return while the safe flight is plotted in Fig 11. Cumulative reward is the moving average of last N rewards received by the agent and is given by $R_i = \frac{1}{N}\sum_{j=i-N}^{i} r_j$ where $i \geq N$ and N is a smoothing constant and was taken to be 15000. The return is the moving average of the sum of rewards across episodes. With each iteration, the agent takes an action and a reward is presented. These rewards are accumulated until the drone crashes and is given by $\frac{1}{N_k}\sum_{j=i-N_k}^{i} r_j$ where $N_k$ is the number of actions taken between the $k^{th}$ and $(k-1)^{th}$ crash. The safe flight [3] is the average distance (in meters) travelled by the drone before it crashes and gives a more quantitative measure of how good the drone is in avoiding obstacles. From Fig. 10 we note that the system converges (saturating reward) for all the three scenarios showing the efficacy of the proposed algorithm. The return shows comparable performance across all the algorithms with $L_4$ showing best performance. The normalized SFD shows acceptable degradation in performance (3% to 8.1%). In outdoor town environments the meta-environment and test environments show large disparities (the type of houses, trees, cars etc. that the drone encounters) and shows the largest degradation. This can be further improved by performing TL on richer meta-environments.

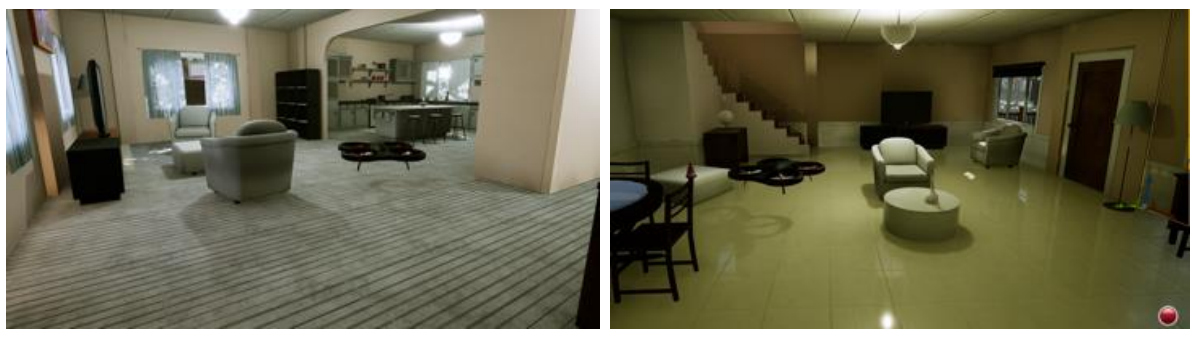

(a) Indoor Apartment (b) Indoor House

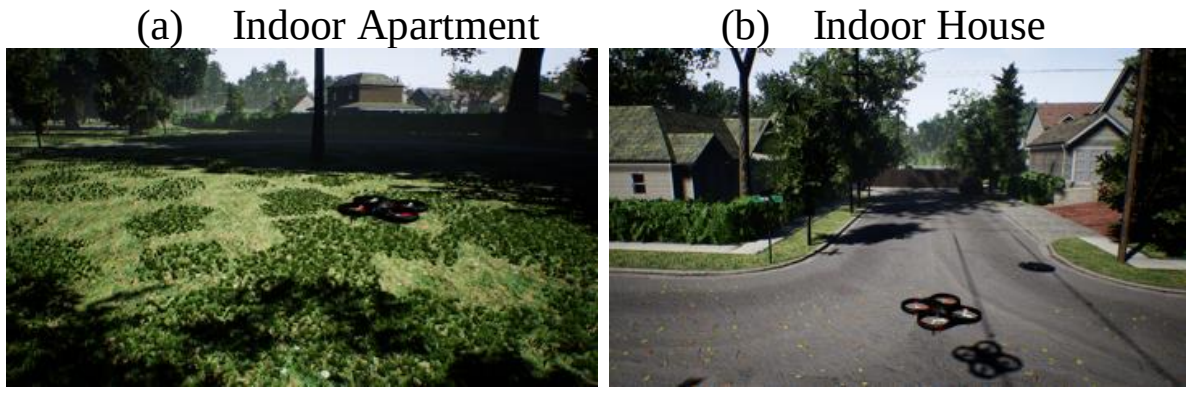

(c) Outdoor Forest (d) Outdoor Town

Fig.9. Typical screenshots of the test environments developed using Unreal Engine 4.

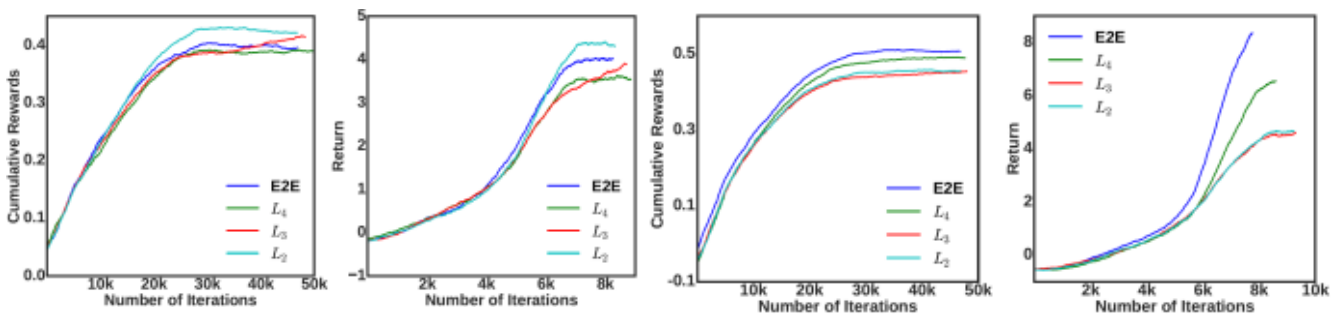

(a) Indoor apartment (b) Indoor house

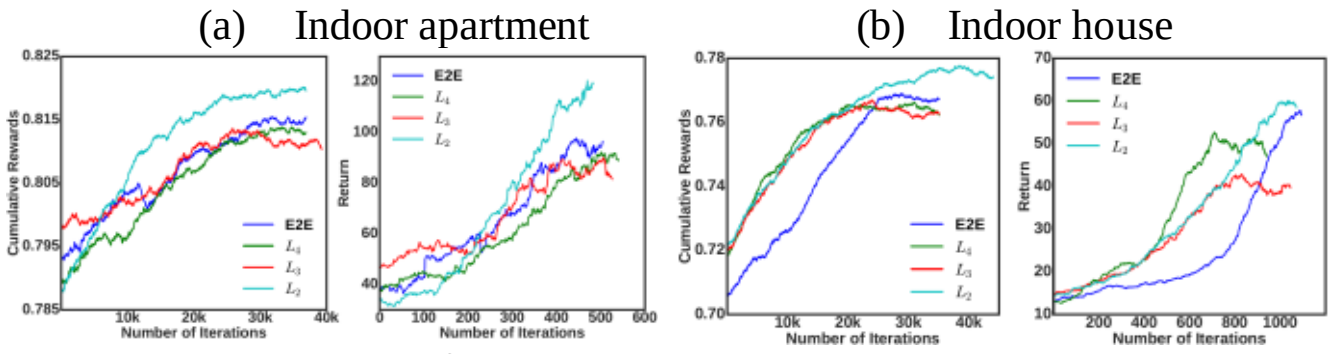

(b) outdoor forest (d) outdoor town

Fig.10. Cumulative rewards and return results in indoor and outdoor test environments. The legend $L_i$ indicates TL with last i-layers. All the algorithms show convergence and improving return loss indicating successful learning.

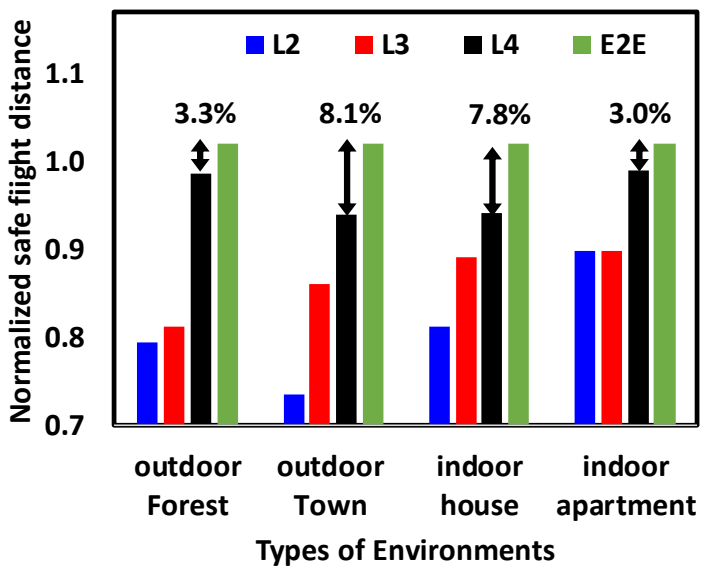

Fig.11 Normalized safe flight distance (SFD) with respect to different environments.

### C. System Evaluation

The hardware system is evaluated and the post-synthesis results are summarized in Fig. 12 and 13. The latency, energy and number of active PEs for the forward and backward propagation of data for each of the layers is shown in Fig. 12. We plot the maximum *fps* that can be supported in the

| Layer | Processing Latency(ms) | Num. of Active PE | Power(mW) | Energy(mJ) |
|---|---|---|---|---|
| CONV1+ReLU+Maxpool | 0.245 | 704 | 4134 | 1.012 |
| CONV2+ReLU+Maxpool | 1.087 | 960 | 5571 | 6.056 |
| CONV3+ReLU | 0.804 | 960 | 5674 | 4.564 |
| CONV4+ReLU | 1.28 | 960 | 5692 | 7.289 |
| CONV5+ReLU+Maxpool | 1.116 | 960 | 5672 | 6.33 |
| FC1+ReLU | 5.365 | 1024 | 6799 | 36.48 |
| FC2+ReLU | 1.189 | 1024 | 6800 | 8.091 |
| FC3+ReLU | 0.562 | 1024 | 6408 | 3.603 |
| FC4+ReLU | 0.28 | 1024 | 6410 | 1.8 |
| FC5+ReLU | 0.0005 | 160 | 1910 | 0.0009 |
| **total** | **11.9285** | **880** | **5507** | **75.2259** |

(a)Forward propagation system results

| Layer | Processing Latency(ms) | Num. of Active PE | Power(mW) | Energy(mJ) | NVM Write |
|---|---|---|---|---|---|
| FC5+ReLU | 0.0027 | 160 | 2094 | 0.006 | No |
| FC4+ReLU | 0.594 | 1024 | 6548 | 3.89 | |
| FC3+ReLU | 1.182 | 1024 | 6162 | 7.284 | |
| FC2+ReLU | 3.839 | 1024 | 5390 | 20.69 | Yes |
| FC1+ReLU | 29.19 | 1024 | 5390 | 157.3 | |
| CONV5+ReLU+Maxpool | 4.661 | 208 | 1888 | 8.804 | |
| CONV4+ReLU | 5.579 | 260 | 2112 | 11.78 | |
| CONV3+ReLU | 4.71 | 260 | 2112 | 9.947 | |
| CONV2+ReLU+Maxpool | 5.518 | 432 | 2850 | 15.73 | |
| CONV1+ReLU+Maxpool | 38.95 | 1024 | 5390 | 209.9 | |
| **total** | **94.2257** | **644** | **3993.6** | **445.331** | |

(b)Backward propagation system results

Fig.12. Latency, power and energy of each layers in forward and backward propagation

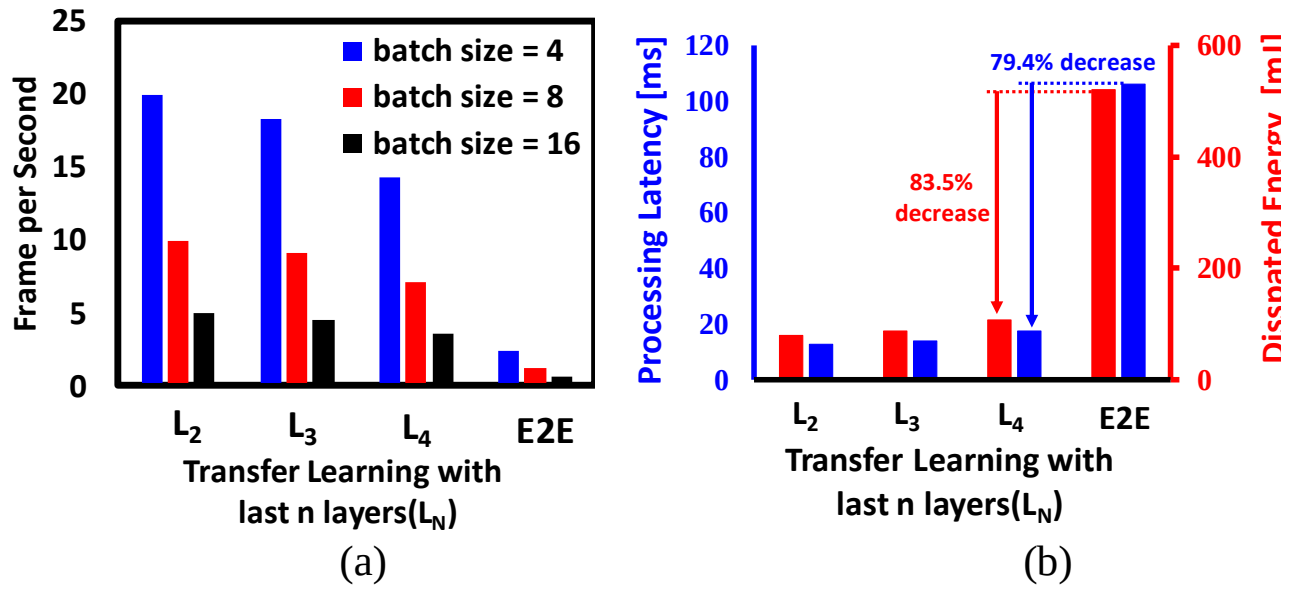

Fig.13: (a) Maximum fps supported by different algorithms as a function of batch size. (b) Estimated processing latency and energy dissipation

proposed system vis-à-vis a baseline E2E RL system. We note that for a batch-size of 4, we can support 15fps for $L_4$, compared to just 3fps for E2E. This directly translates to more than 3X increase in the velocity of the drone (Fig. 1). We also achieve a 79.4% (83.45%) decrease in latency (energy) compared to the baseline. While E2E RL is not feasible in terms of energy and latency for small drones, the proposed solution opens up exciting opportunities for successful autonomous flight under strict power budgets.

## VII. Conclusion

In this paper we present a hardware-algorithm frame-work for STT-MRAM based embedded systems for application to small drones. We show that TL followed by RL on the last few layers of a deep CNN provides comparable performance compared to an E2E RL system, while reducing latency and energy by 79.4% and 83.45% respectively.

## VIII. Acknowledgement

This project was supported by the Semiconductor Research Corporation under grant JUMP CBRIC task ID 2777.006 and JUMP ASCENT task ID 2776.004.